\documentclass[runningheads]{llncs}
\usepackage{graphicx}
\usepackage{booktabs}
\usepackage[table,xcdraw]{xcolor}
\usepackage{microtype}

\begin{document}
\title{Security and Privacy Assessment of U.S. and
Non-U.S. Android E-Commerce Applications}
\titlerunning{Security and Privacy in Global E-Commerce Apps}

\author{
Urvashi Kishnani$^{*}$~\orcidID{0000-0001-6389-5508} \and
Sanchari Das$^{\dagger}$~\orcidID{0000-0003-1299-7867}
}

\institute{
$^{*}$University of Denver, Denver, CO, USA \\
$^{\dagger}$George Mason University, Fairfax, VA, USA
}
\authorrunning{Kishnani and Das}

\maketitle   % typeset the header of the contribution
\begin{abstract}
E-commerce mobile applications are central to global financial transactions, making their security and privacy crucial. In this study, we analyze $92$ top-grossing Android e-commerce apps ($58$ U.S.-based and $34$ international) using MobSF, AndroBugs, and RiskInDroid. Our analysis shows widespread SSL and certificate weaknesses, with approximately $92\%$ using unsecured HTTP connections and an average MobSF security score of $40.92/100$. Over-privileged permissions were identified in $77$ apps. While U.S. apps exhibited fewer manifest, code, and certificate vulnerabilities, both groups showed similar network-related issues. We advocate for the adoption of stronger, standardized, and user-focused security practices across regions.

\keywords{e-commerce applications \and security \and privacy.}
\end{abstract}

\section{Introduction}
The growing demand for convenient and accessible shopping continues to drive the expansion of e-commerce applications~\cite{kishnani2024dual,kishnani2023assessing,kishnani2024securing}. Fellner et al. define e-commerce as ``any business transaction conducted electronically rather than through physical interaction''~\cite{fellner1999component}, while Gaedke and Turowski categorize it into business-to-business (B2B), business-to-consumer (B2C), business-to-administration (B2A), and consumer-to-administration (C2A)~\cite{gaedke2000integrating}. This study focuses on B2C e-commerce through mobile platforms. Understanding regional differences in application security and privacy is crucial for securing user data, ensuring regulatory compliance, and maintaining user trust. Variations in legal frameworks and cybersecurity standards influence data handling practices, making cross-country comparisons essential. While previous studies have explored e-commerce in countries such as Malaysia~\cite{omar2014commerce}, Saudi Arabia~\cite{ahmed2006global}, and India~\cite{chanana2012future}, few have examined global disparities in security and privacy practices~\cite{goyal2019literature,markert2023transcontinental}. This research addresses that gap by answering the following research questions:

\begin{itemize}
    \item \textbf{RQ1:} How does the security robustness of leading e-commerce mobile applications vary by major market region (U.S. vs. non-U.S.)?
    \item \textbf{RQ2:} How do differences in privacy measures reflect the regional privacy regulations governing these applications?
\end{itemize}

In this study, we analyzed $92$ Android e-commerce applications ($58$ U.S.-based and $34$ international) using MobSF, AndroBugs, and RiskInDroid. MobSF and AndroBugs were employed for security evaluation, while MobSF and RiskInDroid were used for privacy assessment. Our main contributions are:

\begin{itemize}
    \item An assessment of $92$ popular e-commerce applications, identifying key vulnerabilities in code, manifest, certificate, and network configurations, as well as privacy risks from excessive or dangerous permissions.
    \item A comparative analysis of U.S. and non-U.S. applications, showing that although overall security and privacy remain weak, U.S. apps performed slightly better across most evaluation metrics.
\end{itemize}

\section{Related Work}
Existing research highlights persistent security and privacy challenges in e-commerce mobile applications, including authentication weaknesses, insecure payment mechanisms, and data breaches~\cite{das2019privacy,ghosh2001software,hussain2013study,hadan2019iot}. These threats span network, application, and insider levels~\cite{ladan2014commerce}, underscoring the importance of strong encryption, secure communication protocols such as SSL/TLS, and regular vulnerability assessments~\cite{yang2008approach}. User education also plays a key role in reducing risks by promoting secure payment practices and timely software updates~\cite{niranjanamurthy2013analysis}. As mobile commerce expands, wireless connectivity and increased data exchange further amplify these security risks~\cite{mai2023mobile,lazara2023security}.

Privacy risks in e-commerce apps arise from extensive user tracking and data sharing with advertisers or analytics services~\cite{choi2020mobile,liu2022privacy,liu2023blockchain}. Prior studies identify issues such as identity tracking, smartphone vulnerabilities, and weak regulatory enforcement~\cite{zhang2013mobile}, emphasizing the need for privacy-preserving architectures and stronger data governance~\cite{liu2023blockchain,bandara2020privacy}. Reducing personal data collection enhances consumer control and trust~\cite{smith2007privacy}, although many users still trade privacy for convenience in location-based services~\cite{ng2002price,poikela2016topic}. Building on these insights, we examine permission usage in leading e-commerce apps to expose security and privacy gaps and propose improvements that strengthen user trust and regulatory compliance.

\section{Method}
To build our dataset, we first compiled a list of leading e-commerce stores and identified their corresponding mobile applications. Two primary data sources were used: AfterShip~\cite{aftership2024topstores} and EcommerceDB. AfterShip lists the top 100 online stores by combining monthly revenue and website traffic, with region-specific filters. We retrieved two lists from AfterShip—one filtered for the United States and another global list without regional restrictions. We also collected the top 50 global e-commerce stores from EcommerceDB, as cited by Acosta-Vargas et al.~\cite{acosta2022accessibility}. These lists were merged and cleaned by removing 84 duplicates. When a store appeared under multiple regions, it was associated with the country holding the largest market share, resulting in 166 unique stores.

\label{sec:method}
\begin{figure}[t]
    \centering
    \includegraphics[width=6cm]{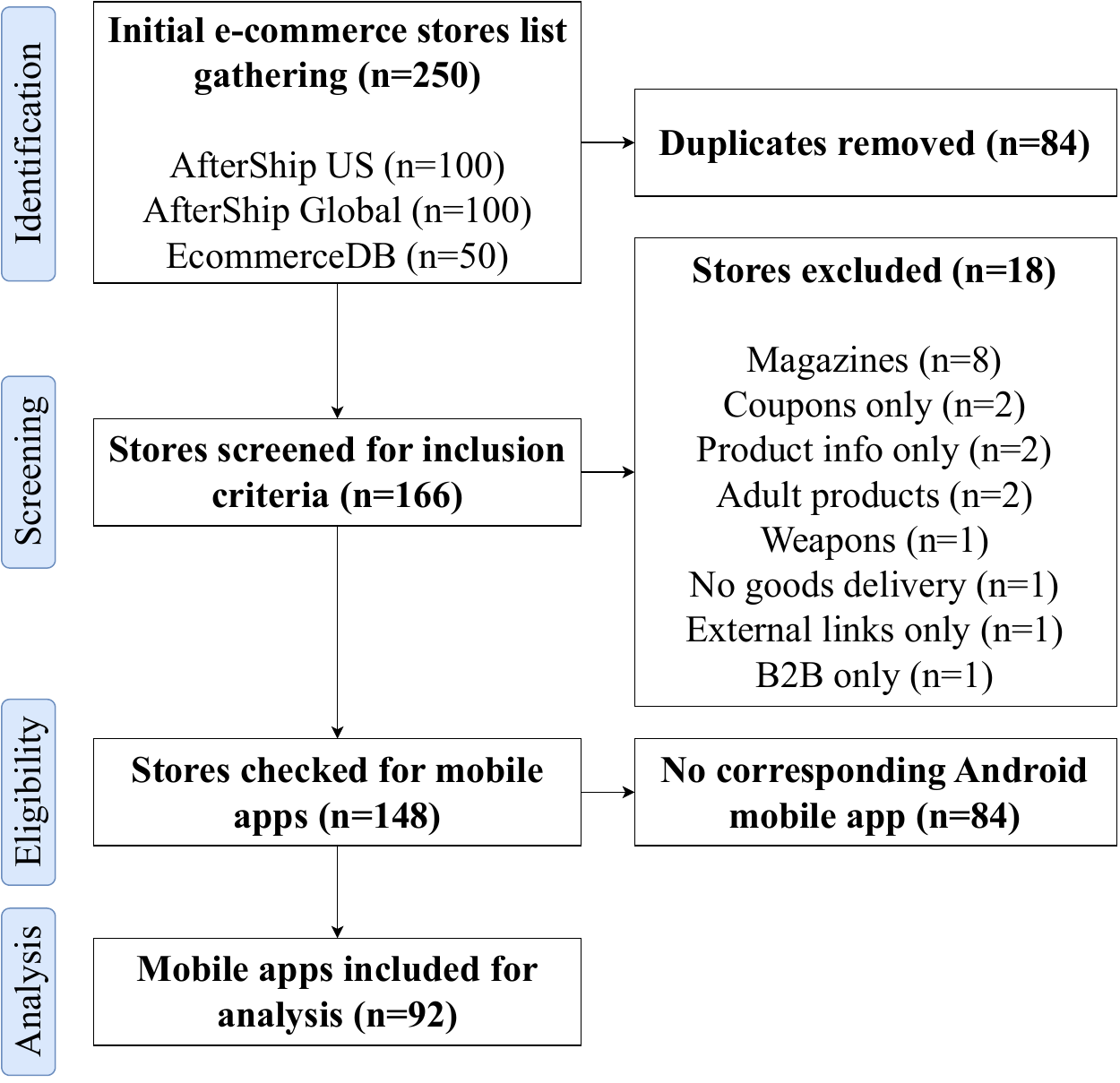}
    \caption{Flowchart of the study method showing the results of the app search and screening process for analysis.}
    \label{fig:methodology}
\end{figure}

\begin{table}[t]
\centering
\caption{Distribution of applications by countries}
\label{tab:app_countries}
\begin{tabular}{lrlrlrlr}
\toprule
\textbf{Country} & \textbf{\#} & \textbf{Country} & \textbf{\#}  & \textbf{Country} & \textbf{\#}  & \textbf{Country} & \textbf{\#} \\ \midrule
United States    & 58  
& Germany            & 3
& Russia           & 2           
& Netherlands          & 1           \\

United Kingdom   & 7           
& India          & 3 
& Sweden           & 2           
& Romania           & 1           \\

Brazil           & 4           
& Japan           & 2 
& Austria       & 1           
& Spain          & 1           \\

China            & 3           
& Poland            & 2
& Mexico             & 1                 
& Thailand          & 1           \\ \bottomrule
\end{tabular}
\end{table}

Eighteen stores were excluded for the following reasons: (i) non-physical goods (digital magazines, coupon-only sites, or product catalogs; $n=13$), (ii) restricted product categories requiring additional privacy and security considerations (adult or weapon-related products; $n=3$), and (iii) B2B-only platforms ($n=2$). From the remaining 148 stores, 92 had corresponding Android applications, which were downloaded in their latest available version as of 2023. Among these, 58 apps targeted the U.S. market, and 34 were international. The geographic distribution of these applications is presented in Table~\ref{tab:app_countries}, and the overall app collection and filtering process is summarized in Figure~\ref{fig:methodology}.

Each application was analyzed using the three tools: MobSF, AndroBugs, and RiskInDroid. MobSF performs static and dynamic analysis of Android, iOS, and Windows applications by reverse engineering APK files~\cite{abraham2016mobile}. It generates detailed reports with severity levels (High, Warning, Info, Secure) and identifies dangerous permissions. AndroBugs conducts static analysis to detect vulnerabilities categorized as Critical, Warning, or Notice~\cite{lin2015androbugs,androbugs2}, while RiskInDroid evaluates privacy-related risks by classifying permissions as Declared, Exploited, Useless, or Ghost~\cite{merlo2017riskindroid}. Reports from all tools were automatically generated and parsed using Python and Bash scripts to ensure consistency and efficiency~\cite{surani2023security,kishnani2023assessing}.

MobSF successfully generated reports for $91$ of $92$ applications, AndroBugs for $76$, and RiskInDroid for $81$, with the remaining apps excluded due to timeouts or large APK sizes. For the security analysis, we compared MobSF and AndroBugs results between U.S. and non-U.S. applications, focusing on high-severity vulnerabilities (Critical and Warning) and average security scores. To maintain consistency, all high-severity findings were labeled as Critical. For the privacy analysis, we used RiskInDroid and MobSF data to assess permission-related risks, including declared, exploited, useless, ghost, and dangerous permissions, along with their average counts and distributions. RiskInDroid analyzed $82$ applications ($51$ U.S., $31$ non-U.S.), and MobSF analyzed $91$ applications ($57$ U.S., $34$ non-U.S.). All APKs were obtained from trusted sources such as Google Play and APKPure. Following ethical guidelines, specific app names are not disclosed, and all identified organizations were informed of the findings, with several confirming that the reported vulnerabilities were resolved.

\section{Results}
\subsection{Security}
We analyzed the security scores and vulnerabilities of all applications, as summarized in Table~\ref{tab:ch3_sec_1}. The table presents the distribution of vulnerabilities by severity (Critical and Warning) across all apps, U.S. apps, and non-U.S. apps, accounting for differences in the number of successfully analyzed applications per tool.

\textbf{Application Security Scores}: MobSF assigns each application a security score ranging from $0$ (lowest) to $100$ (highest), with deductions of $15$ for each Critical and $10$ for each Warning vulnerability. An additional $5$ points are added for exceptional security measures, capped at $100$. Among U.S. apps, scores ranged from $13$ to $67$ with an average of $42$, while non-U.S. apps scored between $16$ and $73$ with an average of $39.12$.

\begin{table*}[htbp]
\caption{SSL and certificate-related, code-related, manifest-related, and network-related vulnerabilities with total, U.S., and non-U.S. averages of critical (C) and warning (W) counts}
\label{tab:ch3_sec_1}
\centering
\begin{tabular}{lrrrrrr}
\toprule
\multicolumn{1}{c}{\textbf{Issue}}                                                              & \multicolumn{2}{c}{\textbf{Total}} & \multicolumn{2}{c}{\textbf{U.S.}} & \multicolumn{2}{c}{\textbf{Non-U.S.}} \\ \midrule
\multicolumn{1}{c}{\textbf{}}                                                                   & \textbf{C}       & \textbf{W}      & \textbf{C}        & \textbf{W}        & \textbf{C}          & \textbf{W}          \\ \midrule
\multicolumn{4}{l}{\textbf{SSL and Certificate-Related Vulnerabilities}} \\ \midrule
HTTP Connection                                                                                 & 92.11 & 0     & 89.36 & 0     & 96.55 & 0     \\
Weak SSL Implementation                                                                         & 32.89 & 0     & 29.79 & 0     & 37.93 & 0     \\
Insecure Implementation of SSL                                                                  & 21.05 & 0     & 17.02 & 0     & 27.59 & 0     \\
Weak SSL Cert. Verification                                                                     & 18.42 & 36.84 & 17.02 & 34.04 & 20.69 & 41.38 \\
Cert. Alg. Vulnerable to Hash Collision                                                         & 12.09 & 34.07 & 12.28 & 33.33 & 11.76 & 35.29 \\
Janus Vulnerability                                                                             & 3.3   & 79.12 & 3.51  & 77.19 & 2.94  & 82.35 \\ \midrule
\multicolumn{4}{l}{\textbf{Code-Related Vulnerabilities}} \\ \midrule
WebView RCE Vulnerability                                                                       & 90.79 & 0     & 89.36 & 0     & 93.1  & 0     \\
CBC with PKCS5/PKCS7 Padding                                                                    & 60.44 & 0     & 56.14 & 0     & 67.65 & 0     \\
Runtime Command                                                                                 & 56.58 & 0     & 51.06 & 0     & 65.52 & 0     \\
Strandhogg 2.0                                                                                  & 55.26 & 0     & 51.06 & 0     & 62.07 & 0     \\
Weak KeyStore Protection                                                                        & 44.74 & 0     & 42.55 & 0     & 48.28 & 0     \\
Remote WebView Debugging                                                                        & 41.76 & 0     & 43.86 & 0     & 38.24 & 0     \\
Implicit Service                                                                                & 36.84 & 0     & 27.66 & 0     & 51.72 & 0     \\
Base64 String Encoding                                                                          & 30.26 & 0     & 31.91 & 0     & 27.59 & 0     \\
Fragment Vulnerability                                                                          & 27.63 & 0     & 25.53 & 0     & 31.03 & 0     \\
Insecure WebView   Implementation                                                               & 27.47 & 0     & 24.56 & 0     & 32.35 & 0     \\
ECB-Mode in Cryptography                                                                        & 18.68 & 0     & 17.54 & 0     & 20.59 & 0     \\
Weak AES ECB Mode                                                                               & 9.89  & 0     & 10.53 & 0     & 8.82  & 0     \\
Weak Encryption Algorithm Used                                                                  & 6.59  & 0     & 1.75  & 0     & 14.71 & 0     \\
App Sandbox Permission   Checking                                                               & 3.95  & 0     & 0     & 0     & 10.34 & 0     \\
Runtime Critical Command                                                                        & 3.95  & 0     & 0     & 0     & 10.34 & 0     \\
World Readable File                                                                             & 1.1   & 0     & 0     & 0     & 2.94  & 0     \\
Weak Cryptographic Algorithms                                                                   & 1.1   & 0     & 1.75  & 0     & 0     & 0     \\ \midrule        
\multicolumn{4}{l}{\textbf{Manifest-Related Vulnerabilities}} \\ \midrule
Activity is not Protected                                                                       & 69.23 & 0     & 64.91 & 0     & 76.47 & 0     \\
Service is not Protected                                                                        & 63.74 & 0     & 57.89 & 0     & 73.53 & 0     \\
Broadcast Receiver is not Protected                                                             & 58.24 & 0     & 54.39 & 0     & 64.71 & 0     \\
Clear text traffic is Enabled For App                                                           & 19.78 & 0     & 17.54 & 0     & 23.53 & 0     \\
ContentProvider  is not Protected                                                               & 14.29 & 0     & 10.53 & 0     & 20.59 & 0     \\
ContentProvider Exported                                                                        & 11.84 & 0     & 14.89 & 0     & 6.9   & 0     \\
Launch Mode of Activity  is not standard                                                        & 9.89  & 0     & 7.02  & 0     & 14.71 & 0     \\
Activity-Alias  is not Protected                                                                & 7.69  & 0     & 7.02  & 0     & 8.82  & 0     \\
Intent Filter Settings                                                                          & 3.95  & 0     & 4.26  & 0     & 3.45  & 0     \\ \midrule
\multicolumn{4}{l}{\textbf{Network-Related Vulnerabilities}} \\ \midrule
Domain Conf. Permitting Clear-text Traffic                                                      & 17.58 & 0     & 19.3  & 0     & 14.71 & 0     \\
Base Conf. Permitting Clear-text Traffic                                                        & 16.48 & 0     & 17.54 & 0     & 14.71 & 0     \\
Base Conf. Trusting User Installed Certs.                                                       & 3.3   & 0     & 1.75  & 0     & 5.88  & 0     \\
Domain Conf. Trusting User Installed Certs.                                                     & 1.1   & 0     & 1.75  & 0     & 0     & 0     \\
Base Conf. Bypassing Certificate Pinning                                                        & 1.1   & 0     & 0     & 0     & 2.94  & 0  \\ \bottomrule         
\end{tabular}
\end{table*}

Proper implementation of \textbf{Secure Socket Layer (SSL)} and \textbf{Transport Layer Security (TLS)} is essential for secure communication and certificate verification. However, about $92\%$ of the applications used unsecured HTTP connections, and nearly one-third failed to validate hostnames or certificates correctly. Around $20\%$ trusted all or self-signed certificates, making them susceptible to Man-in-the-Middle (MITM) attacks, while over half did not verify certificate validity or CN-field consistency. Approximately $12\%$ of apps still used the outdated SHA1withRSA algorithm, exposing them to collision attacks. Some applications also exhibited the Janus vulnerability, where malicious DEX files can be injected into APKs with $v1$ signature schemes on Android versions $5.0$–$8.0$, posing a critical risk.

\textbf{Code-related weaknesses} were widespread due to outdated practices and insecure use of Android components. About $90\%$ of apps contained the WebView Remote Code Execution vulnerability (CVE-2013-4710), allowing Denial of Service and arbitrary code execution. Additionally, $45\%$ permitted remote debugging, and $27\%$ ignored SSL certificate errors, increasing MITM exposure. Roughly $37\%$ of apps had implicit service vulnerabilities, and $27\%$ were affected by the Fragment vulnerability (CVE-2013-6271), enabling unauthorized control. Some apps also misused deprecated permissions (\texttt{MODE\_WORLD\_READABLE}/\texttt{WRITEABLE}) that compromise sandboxing. More than half the apps were vulnerable to Strandhogg $2.0$, which allows malicious overlays on legitimate apps to steal sensitive data. Runtime command injection was also common, with some apps executing commands through \texttt{Runtime.getRuntime().exec()}, including a few attempting privileged “su” execution. Around $60\%$ were susceptible to padding oracle attacks due to improper CBC mode and padding usage, and $45\%$ showed weak keystore protection through hardcoded keys. Several apps also used weak or outdated cryptography (ECB and AES-ECB modes) or insecure Base64 encoding. 

The Android \textbf{Manifest} defines app permissions, components, and compatibility settings~\cite{android_developers_guide}. Misconfigurations in this file can expose components such as \texttt{Activity}, \texttt{Service}, \texttt{BroadcastReceiver}, and \texttt{ContentProvider}. About $70\%$ of apps had unprotected activities, $63\%$ shared services, and $58\%$ insecure broadcast receivers, allowing unauthorized access by other applications. ContentProvider misconfigurations were present in $14\%$ of apps, often leading to data leakage, while $12\%$ relied on default exported settings, further increasing exposure. Around $20\%$ of apps transmitted cleartext data, compromising confidentiality and integrity, and $10\%$ misused restricted launch modes (\texttt{singleTask}, \texttt{singleInstance}), risking data leaks. A few apps also had improperly configured intent filters that exposed sensitive components to unintended access~\cite{maji2012empirical,chin2011analyzing}.

\textbf{Network-Related Vulnerabilities} were primarily linked to weak security configurations in the Manifest’s network settings file~\cite{android_developers_guide}. Several apps permitted cleartext traffic, trusted user-installed certificates, or bypassed certificate pinning, leaving them vulnerable to data interception and tampering. The most prevalent issue across both base and domain configurations was allowing cleartext transmission, potentially exposing sensitive information.

\subsection{Privacy}
We examined requested, used, and tracked permissions, reported as percentages due to differing tool coverage (Table~\ref{tab:ch3_privacy}). Across all applications, $157$ unique permissions were declared, $146$ of which were unused, totaling $1,169$ instances. All but four apps requested at least one dangerous permission.

\textbf{Permission-Based Risk Scores}: 
RiskInDroid assigns each app a score from $0$ to $100$, representing the likelihood of malicious behavior, where lower scores indicate lower risk. Among U.S. apps, scores ranged from $6.13$ to $46.87$, while non-U.S. apps ranged from $10.84$ to $72.92$.

\begin{table*}[t]
\caption{Privacy metrics with total, U.S., and Non-U.S. averages}
\label{tab:ch3_privacy}
\centering
\begin{tabular}{lrrr}
\toprule
\textbf{Privacy   Metric}                     & \textbf{Total} & \textbf{U.S.} & \textbf{Non-U.S.} \\ \midrule
Average Permission-based Risk Score           & 22.57          & 20.44             & 26.18                 \\ 
Best (Lowest) Permission-based Risk Score     & 6.13           & 6.13              & 10.84                 \\ 
Worst (Highest) Permission-based Risk   Score & 72.92          & 46.87             & 72.92                 \\ \midrule
Average Number of Declared Permissions                  & 21.58          & 20.22             & 23.9                  \\ 
Average Number of Exploited Permissions                 & 7.15           & 7.02              & 7.37                  \\ 
Average Number of Useless Permissions                   & 14.46          & 13.24             & 16.53                 \\ \midrule
Average Number of Ghost Permissions                     & 7.74           & 8.1               & 7.13                  \\ \midrule
Average Number of Dangerous Permissions                 & 5.8            & 5.37              & 6.53                  \\ \bottomrule
\end{tabular}
\end{table*}
\textbf{Declared Permissions} appear in the app Manifest but may not always be used in the code. The most common declared permissions were \texttt{INTERNET} ($80$), \texttt{ACCESS\_NETWORK\_STATE} ($79$), \texttt{RECEIVE} ($77$), \texttt{WAKE\_LOCK} ($76$), and \texttt{BIND\_GET\_INSTALL\_REFERRER\_SERVICE} ($73$). Notably, several declared permissions were unused, leading to over-privileged apps that can raise compliance and transparency concerns.

\textbf{Ghost Permissions} are invoked in code but not declared in the Manifest, often due to third-party libraries. Such inconsistencies can lead to runtime failures or unauthorized access attempts. The most frequent ghost permissions were \texttt{READ\_PROFILE} ($77$), \texttt{MANAGE\_ACCOUNTS} ($65$), \texttt{WRITE\_SETTINGS} ($60$), \texttt{AUTHENTICATE\_ACCOUNTS} ($54$), and \texttt{GET\_ACCOUNTS} ($48$).

\textbf{Dangerous Permissions} can access sensitive user data and require explicit user consent at runtime. The most frequent were \texttt{CAMERA} ($76$), \texttt{ACCESS\_FINE\_LOCATION} ($65$), \texttt{WRITE\_EXTERNAL\_STORAGE} ($64$), \texttt{ACCESS\_COARSE\_LOCATION} ($57$), and \texttt{READ\_EXTERNAL\_STORAGE} ($56$). While these support legitimate app features such as AR views, file uploads, or location-based services, improper use may expose users to privacy and security risks.

\textbf{Privacy Trackers} collect user data for analytics or advertising~\cite{monogios2022privacy}. Among the $91$ apps analyzed with MobSF, $64$ contained at least one tracker. Undisclosed tracking practices raise privacy concerns and erode user trust~\cite{boritz2011commerce,bandara2020privacy}. Developers should clearly disclose tracking activities, explain their purpose, and provide users with the option to opt out to ensure transparency and regulatory compliance.

\subsection{Discussion}

\textit{RQ1:} Our analysis revealed that e-commerce applications, regardless of region, showed notable security weaknesses, with an average score of $40.92$ out of $100$. Although some vulnerabilities, such as $CVE$-$2013$-$4710$ and $CVE$-$2013$-$6271$, affect older Android versions used by less than $0.5\%$ of users, they still pose risks to millions worldwide~\cite{businessofapps2024androidstats}. U.S. apps achieved slightly higher security scores (maximum $67$) compared to non-U.S. apps (maximum $73$), with fewer critical vulnerabilities overall. These gaps mainly stem from weak manifest, certificate, and network configurations, underscoring the need for stronger secure coding practices, encryption, and vulnerability mitigation. The comparatively better performance of U.S. apps may reflect stricter development standards and larger-scale operations enforcing tighter controls.

\textit{RQ2:} Privacy analysis showed that U.S. applications generally demonstrated stronger privacy practices, including lower permission-based risk scores, fewer unnecessary permissions, and reduced reliance on dangerous ones. These results align with the influence of regulations such as the CCPA and GDPR, where GDPR enforces explicit security obligations while CCPA focuses on data privacy rights. Non-U.S. apps, especially those popular in EU regions, exhibited more ghost permissions, indicating inconsistent permission declarations and possible overreach in data collection. Our analysis emphasize the need for continuous improvement in transparency, permission management, and compliance mechanisms to strengthen user trust and safeguard sensitive e-commerce data globally.

\section{Limitations and Future Work}
We analyzed the security and privacy of U.S. and non-U.S. e-commerce mobile applications using MobSF, RiskInDroid, and AndroBugs. Although some tools failed on a few apps due to technical constraints, we ensured that each app was evaluated by at least one tool. In future work, we will incorporate dynamic analysis to capture runtime behavior, expand our scope to include e-payment, banking, and iOS applications, and examine how security and privacy measures influence usability and user experience.

\section{Conclusion}
\label{conclusion}
E-commerce applications serve over $2.5$ billion users worldwide, making strong security and privacy measures vital for protecting sensitive data and maintaining user trust. Yet, many studies overlook regional differences in app security and privacy. In this work, we analyzed $92$ high-revenue Android e-commerce applications from the United States ($58$) and other countries ($34$) using three open-source tools: MobSF, RiskInDroid, and AndroBugs. Our analysis revealed critical vulnerabilities such as remote WebView debugging flaws that enable DDoS and code injection attacks, and privacy trackers found in nearly $70\%$ of apps, indicating weak privacy controls. We recommend that regulators enforce stricter compliance and audits, and that developers apply secure coding, continuous testing, and privacy-by-design practices to strengthen global e-commerce app security.

\bibliographystyle{splncs04}

\bibliography{web}
\end{document}